\documentclass[proceedings, preprint]{rmaa}

\usepackage{paralist}

\usepackage{psfrag,color}

\SetYear{2012}
\SetConfTitle{Science with the GTC: IV Meeting}

\title{A mid-IR survey of local AGN with GTC/CanariCam}

\author{A. Alonso-Herrero\altaffilmark{1,2}
  and C. Packham\altaffilmark{3}}

\altaffiltext{1}{Instituto de F\'{\i}sica de Cantabria, CSIC-UC, 39005
  Santander, Spain (aalonso@ifca.unican.es).} 

\altaffiltext{2}{Augusto Gonz\'alez Linares Senior Research Fellow.}

\altaffiltext{3}{Department of Astronomy, University of Florida,
  Gainesville, FL 32611, USA (packham@astro.ufl.edu)}

\shortauthor{Alonso-Herrero \& Packham}
\shorttitle{A mid-IR survey of local AGN with GTC/CanariCam}

\listofauthors{A. Alonso-Herrero, C. Packham}
\indexauthor{Alonso-Herrero, A., Packham, C.}
\abstract{We describe a mid-infrared (MIR) survey of local AGN to be
  conducted with the CanariCam instrument on the Gran Telescopio
  Canarias (GTC). We will obtain MIR imaging and spectroscopy of a
  sample of $\sim 100$ AGN covering six orders of magnitude in
  AGN luminosity, and including different AGN classes (e.g., LINERs, Seyfert
1s and 2s, QSO). The main goals are: (1) to test unification of Type 1
and Type 2 AGN, (2) to study the star formation activity around AGN
and (3) to explore the role of the dusty torus in low-luminosity AGN. }

\resumen{Presentamos una exploraci\'on en el infrarrojo medio (MIR)
de una muestra de AGN cercanos usando CanariCam en el Gran Telescopio
Canarias (GTC). Se van a obtener im\'agenes y espectroscopia de unos
100 AGN   cubriendo seis \'ordenes de magnitud en luminosidad del AGN
y diferentes tipos (p.e., LINERs, Seyfert
1s y 2s, QSO). Los objetivos principales son: (1) testear el modelo
unificado para AGN de Tipos 1 y Tipos 2, (2) estudiar la formaci\'on
estelar alrededor de AGN y (3) explorar el papel del toro de polvo en
AGN de baja luminosidad.}

\addkeyword{Galaxies: AGN}
\addkeyword{Galaxies: infrared}
\addkeyword{Galaxies: Star Formation}
\begin{document}
% Typeset article header
\maketitle

\section{Introduction}
\label{sec:intro}
Active galactic nuclei (AGN) are powered by accretion of gas  
onto a supermassive ($10^6-10^9\,{\rm M}_\odot$) black hole. 
The Unified Model \citep{Antonucci1993} explains the observed
differences between AGN classified as Type 1 and Type 2---notably the
presence or absence of broad optical emission lines---in terms of
geometry.  The central engines of Type 1 AGN are viewed directly.  In
the Type 2 views, an optically and geometrically thick "torus" of
material in the inner region hides the active nucleus and broad
emission-line region.  Moreover, unification is extended to encompass
both radio quiet (RQ) and radio loud (RL) objects, which exhibit
either a low or high ratio of optical to radio luminosity,
respectively \citep{Urry1995}. 

Whilst critical to the Unified Model of AGN, the torus itself
remains poorly constrained.  Since torus models agree that the  torus peaks at
mid-infrared (MIR, $\sim 8-26\,\mu$m) wavelengths, MIR observations
offer the best opportunity for progress in 
constraining the torus models. Although the pc-scale torus can
only be resolved with interferometric observations, this  technique
can only be used for a few very bright and nearby sources \citep{Tristram2009}, 
definitely not enough for any statistical investigation.
The MIR angular resolution of 8-10m class telescopes although is not
sufficiently high to spatially resolve the torus structure, it can help
 disentangle the  torus emission from the 
circumnuclear and host galaxy contribution, as well as from  the emission 
of the narrow line region \citep{Radomski2003,Packham2005,Roche2006}. 
This is most relevant for
modeling the properties of the 
torus, as in the past the use of MIR large aperture
measurements resulted in a confused 
picture of
the torus, accretion, and interaction with the host galaxy.

\section{The Survey}
\label{sec:survey}
We will be carrying out a MIR survey of a sample of local AGN using
the CanariCam \citep{Telesco2003} instrument 
on the Gran Telescopio Canarias (GTC). The
observations will be taken as part of the guaranteed time of the
CanariCam AGN Science Team and an ESO/GTC large programme. We used the hard ($2-10\,$keV) 
X-ray luminosity as a proxy for the AGN power, to select a sample 
spanning AGN from the
lowest activity (e.g., LINERs) to the highest activity  
(e.g. quasars).
The sample includes about 100 AGN, and 
covers more than six decades in 2-10\,keV X-ray luminosity 
($\sim 3 \times 10^{38}$ to $\sim 3 \times 
10^{45}\,{\rm erg \, s}^{-1}$). To
construct the full sample, we balanced 
the X-ray bins in: (1) X-ray luminosity, and (2) 
AGN type: RL/RQ, Type1/2,
LINER, with/without star formation (SF). We will observe the galaxies
with the imaging mode using two filters, namely the Si-2
($\lambda_{\rm c}=8.7\,\mu$m) filter and the Q4 filter ($\lambda_{\rm
  c}=20.5\,\mu$m), and  with the low-resolution spectroscopic mode
covering the $\sim 
8-13\,\mu$m spectral range. In addition we will obtain polarimetric
observations for a few of the brightest AGN in our sample.

\begin{figure}[!t]
  \includegraphics[angle=-90,width=7.8cm]{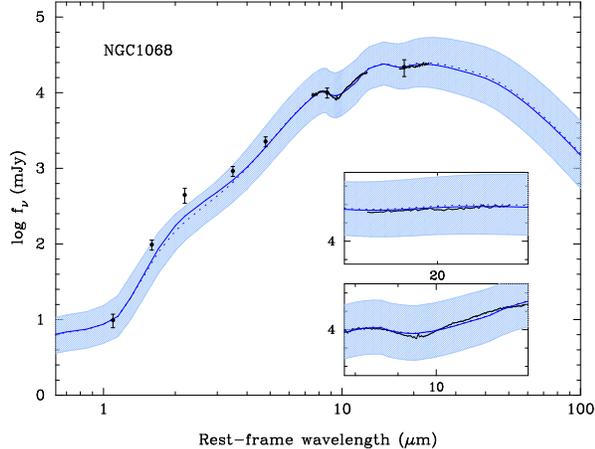}%
  \caption{An example of the fit to the nuclear near-infrared and MIR
    photometric points, and MIR spectroscopy of 
NGC~1068 using the \citet{Nenkova2008} clumpy torus
    models. The insets show in detail the spectral regions around
    the $10\,\mu$m and $18\,\mu$m silicate features.  Figure 
adapted from \citet{AlonsoHerrero2011}.}
  \label{fig:torusmodelfits}
\end{figure}

\section{Goals of the Survey}
\label{sec:goals}
We describe here some of the main goals of our survey and 
discuss some scientific results obtained with similar
instruments on other 8-10m class telescopes, albeit for much smaller
samples. These should serve as illustrations of the
potential  of our CanariCam survey.
\subsection{Unification of Type 1 and Type 2 AGN}
There has been claims in the literature of a  luminosity dependence of the
Type1/Type2 ratio of AGN, with the fraction of Type 1 objects increasing for
higher luminosities \citep{Simpson2005}. The 
physical origin of this relation is unclear, and several authors
proposed a change in the geometry of the obscuring
material around AGN.  Through the comparison of the MIR imaging and
spectroscopic  CanariCam observations  and models,
spanning a wide AGN luminosity range, we will check whether the decrease
of obscuration as a function of luminosity is due to a change of the
 distribution of the clouds, or of their number, or of
their individual optical thickness.

Fig.~\ref{fig:torusmodelfits} shows  an example of the successful
application of the  
\citet{Nenkova2008} clumpy torus models to fit the nuclear infrared
emission, including MIR spectroscopy, of NGC~1068. The MIR
data have obtained with other 8-10m class telescopes with similar
instrumentation \citep[see][for full details]{AlonsoHerrero2011}.
This kind of model fitting is particularly useful not only to estimate
the torus model parameters, but also to recover the intrinsic AGN
luminosity. Our recent work with 
small samples seems to
indicate some dependency of the torus geometry on the Type1/Type2 class
\citep{RamosAlmeida2011} 
 and/or luminosity of the AGN \citep{AlonsoHerrero2011}. We will
 perform a similar modelling for a more statistically significant sample of
  AGN using the CanariCam and complementary observations.

\begin{figure}[!t]
\hspace{0.5cm}
  \includegraphics[width=7cm]{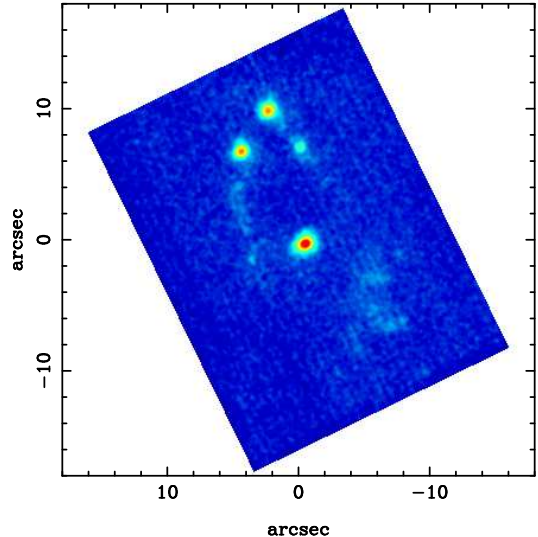}%
  \caption{Gemini/T-ReCS Qa-band ($\lambda_{\rm c}=18.3\,\mu$m) 
image of the central region of the Seyfert
    1.8 galaxy NGC~1365. Adapted from \citet{AlonsoHerrero2012}.}
  \label{fig:ngc1365}
\end{figure}

 \begin{figure*}[!t]
\hspace{1cm}
  \includegraphics[width=14.7cm]{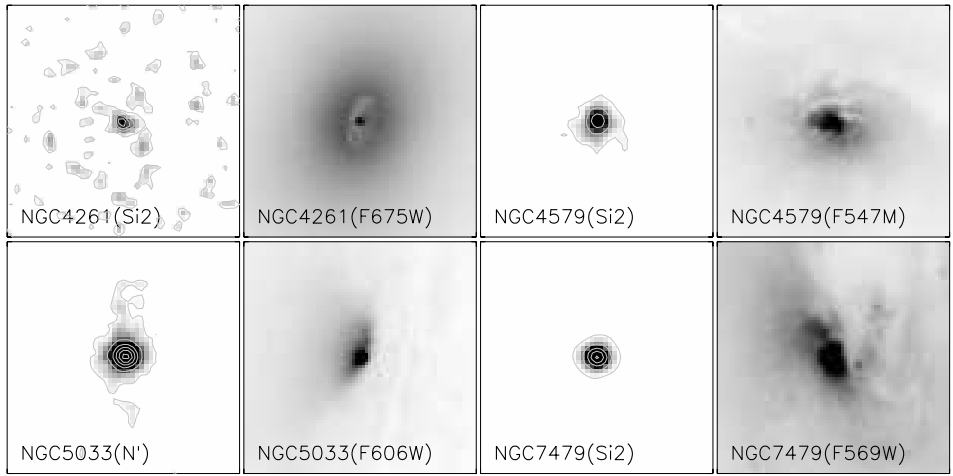}%
\vspace{-0.7cm}
  \caption{Comparison between Gemini/T-ReCS or Michelle MIR (N' and
    Si-2 filters) 
     and HST optical (F547M, F569W, F606W, F675W filters) 
morphologies of the
    central regions of 4 LLAGN. 
Figure adapted from \citet{Mason2012}.}
  \label{fig:LINERs}
\end{figure*}

\subsection{Star formation activity around AGN}
Since material must 
be driven inwards from the ISM of the host galaxy to the nucleus ($<10\,$pc
regions)   to fuel the AGN, nuclear      
SF appears to be an inevitable consequence of this process. 
Although there is evidence of the presence of SF in AGN, especially in
Type 2s, there are also a number of caveats. Classical SF indicators 
(UV, H$\alpha$, near-IR emission lines) are difficult to use in Type 1 AGN, as
they can be easily contaminated by bright
AGN emission.   It is then not clear if there is  increased
SF activity in the hosts of Type 2 AGN
\citep{Maiolino1995} and whether the SF level 
depends on the activity
class \citep[e.g., radio galaxy, QSO, see][]{Shi2007} and/or AGN luminosity
and Type.  The limited angular resolution of most of the previous  work
only probed regions on kpc scales but with CanariCam we will be able to resolve
smaller physical scales.

We will use the  spatially-resolved CanariCam observations 
to trace the SF activity around AGN, in particular 
the extent and strength of nuclear  and circumnuclear 
star-forming regions in AGN.  We have already used 
Gemini/T-ReCS imaging and spectroscopy to study the spatial
distribution of the nuclear and circumnuclear activity of a small
number of galaxies hosting a Seyfert nuclei
\citep{AlonsoHerrero2006, DiazSantos2008, DiazSantos2010}. In
Fig.~\ref{fig:ngc1365} we show a Gemini/T-ReCS $18.3\,\mu$m image 
of the nearby ($d=18.6\,$Mpc) Seyfert 1.8 galaxy NGC~1365.
In the case of this 
galaxy, there is some extended MIR emission around the AGN 
(inner $\sim 40\,$pc), but most
of the (obscured) 
SF activity in the central regions, as probed by the $18.3\,\mu$m
emission, is  in a $\sim 2\,$kpc-diameter ring
\citep[see][]{AlonsoHerrero2012}.

\subsection{LLAGN and the origin of the torus} 
LINERs are detected in the nuclei of about 50\% of  nearby 
spirals and ellipticals  \citep{Ho2008}, and there is an on-going
debate on whether they are powered by low-luminosity AGN (LLAGN) and/or SF. 
The detection of broad H$\alpha$ components means that at least, some
LINERs do contain an AGN.
On the other hand, 
the possible lack of tori is predicted by the disk wind model which
states that the torus originates as an accretion disk wind.
Fortunately, this  model makes a testable
prediction: below $L_{\rm bol} \sim 10^{42}\,{\rm erg\,s}^{-1}$  mass
accretion can no longer sustain the outflow 
required for large obscuring columns 
\citep{Elitzur2006}. 
Moreover, how the properties of LLAGN tori, if they exist, might 
relate to those in higher luminosity AGN is completely open to question.

We will use the  CanariCam data  
to search for thermal dust emission from the AGN torus. In our recent work
using the MIR instruments on the Gemini telescopes  
we have found a variety of MIR morphologies in a sample of
LLAGN \citep{Mason2012}. Sources with higher Eddington
ratios tend to show more compact and bright MIR sources
(see a few examples in Fig.~\ref{fig:LINERs}), whereas
LLAGN with low Eddington ratios appear more diffuse and
extended in the MIR.  It is
indeed in the former type of objects in which the torus is expected to be
detectable in our CanariCam survey. This is because  
mechanisms (i.e., synchrotron emission,
nuclear SF) other than the dusty torus may contribute
significantly to the nuclear MIR emission of LLAGN.

We thank the CanariCam AGN Science Team for their support
and especially R. Mason and N. Levenson for sharing some of their
data prior to publication.


\begin{thebibliography}
\bibitem[Alonso-Herrero et
  al.(2012)]{AlonsoHerrero2012}Alonso-Herrero, A. et al.  2012, \mnras,
  submitted
\bibitem[Alonso-Herrero et
  al.(2011)]{AlonsoHerrero2011}Alonso-Herrero, A. et al.  2011, \apj,
  736, 82
\bibitem[Alonso-Herrero et
  al.(2006)]{AlonsoHerrero2006}Alonso-Herrero, A. et al.  2006, \apj,
  652, L83
\bibitem[Antonucci(1993)]{Antonucci1993} Antonucci, R. 1993, \araa,
  31, 473
\bibitem[D\'{\i}az-Santos et al.(2010)]{DiazSantos2010}
  D\'{\i}az-Santos, T. et al. 2010, \apj, 711, 328
\bibitem[D\'{\i}az-Santos et al.(2008)]{DiazSantos2008}
  D\'{\i}az-Santos, T. et al. 2008, \apj, 685, 211
\bibitem[Elitzur \& Shlosman(2006)]{Elitzur2006} Elitzur, M., \&
  Shlosman, I. 2006, \apj, 648, L101
\bibitem[Ho (2008)]{Ho2008} Ho, L. C. 2008, \araa, 46, 475
\bibitem[Maiolino \& Rieke(1995)]{Maiolino1995} Maiolino, R., \&
  Rieke, G. H. 1995, \apj, 445, 561
\bibitem[Mason et al.(2012)]{Mason2012} Mason, R. et al. 2012
  \apj, in press (astro-ph/1205.0029)
\bibitem[Nenkova et al.(2008)]{Nenkova2008} Nenkova, M. et al.2008,
  \apj, 685, 160
\bibitem[Packham et al. (2005)]{Packham2005} Packham, C., et al. 2005, \apj, 618, L17
\bibitem[Radomski et al. (2003)]{Radomski2003} Radomski, J. T., et al. 2003, \apj, 587, 117
\bibitem[Ramos Almeida et
  al.(2011)]{RamosAlmeida2011}Ramos Almeida, C. et al. 2011, \apj,
  731, 92
%\bibitem[Ramos Almeida et
%  al.(2009)]{RamosAlmeida2009}Ramos Almeida, C. et al. 2009, \apj,
%  702, 1127
\bibitem[Roche et al. (2006)]{Roche2006} Roche, P. F. et al. 2006, \mnras, 367, 1689
\bibitem[Shi et al. (2007)]{Shi2007} Shi, Y. et al. 2007, \apj, 669,
84
\bibitem[Simpson(2005)]{Simpson2005} Simpson, C. 2005, \mnras, 360, 565
\bibitem[Telesco et al. (2003)]{Telesco2003} Telesco, C. M. et
  al. 2003, SPIE, 4841, 913
\bibitem[Tristram et al. (2009)]{Tristram2009} Tristram, K. R. W. et  al. 2009, \aap, 502, 67
\bibitem[Urry \& Padovani(1995)]{Urry1995} Urry, C.M., \& Padovani, P. 1995, \pasp, 107, 803 

\end{thebibliography}
\end{document}